%% file: 000-main.tex
\documentclass[10pt,conference]{IEEEtran}
\IEEEoverridecommandlockouts

\usepackage{cite}
\usepackage{amsmath,amssymb,amsfonts}
\usepackage{algorithmic}
\usepackage{graphicx}
\usepackage{textcomp}
\usepackage{xcolor}
\usepackage{hyperref}
\usepackage{listings}
\usepackage[ruled, linesnumbered]{algorithm2e}
\usepackage{lipsum}
\usepackage{multirow}
\usepackage{subcaption}
\usepackage{booktabs}
\usepackage{threeparttable}
\usepackage{url}
\usepackage{fancyhdr}
\usepackage{indentfirst}
\usepackage{colortbl}
\usepackage{dblfloatfix}
\usepackage{makecell}
\usepackage{array}
\usepackage{enumitem}
\usepackage{hhline}
\usepackage{float}
\usepackage[binary-units]{siunitx}
\usepackage[export]{adjustbox}
\usepackage{soul}
\usepackage{pifont}
\usepackage{tcolorbox}
\usepackage{comment}
\usepackage{xspace}
\usepackage{tabularx}

\input{macros.tex}
\begin{document}
\pagenumbering{arabic}
\pagestyle{plain}

\title{Design Space Exploration of DMA based Finer-Grain Compute Communication Overlap}

\author{
  Shagnik Pal$^{1,2}$, 
  Shaizeen Aga$^{1}$,
  Suchita Pati$^{1}$, 
  Mahzabeen Islam$^{1}$,
  Lizy K. John$^{2}$ \\
  $^{1}$Advanced Micro Devices Inc. ,
  $^{2}$The University of Texas at Austin\\
  shagnik@utexas.edu, \{shaizeen.aga, suchita.pati, mahzabeen.islam\}@amd.com, ljohn@ece.utexas.edu
}

\maketitle
\thispagestyle{plain}
\begin{abstract}
\input{001-abstract}

\end{abstract}


\input{002-intro}
\input{003-background}
\input{004-motivation}
\input{005-characterization}
\input{006-ficco_dse}
\input{008-eval}
\input{009-related_works}
\input{010-discussion}
\input{011-conclusion}

\bibliographystyle{IEEEtranS}
\bibliography{references.bib}
\end{document}

%% file: macros.tex
\long\def\ignore#1{}
\sloppypar

\usepackage{soul}

\newcommand{\remove}[1]{}

\newcommand{\tickmark}{\textcolor{green!60!black}{\ding{51}}}
\newcommand{\crossmark}{\textcolor{red}{\ding{55}}}

\newcommand{\dil}{\mbox{DIL}\xspace}
\newcommand{\cil}{\mbox{CIL}\xspace}
\newcommand{\dilfull}{\mbox{Decomposition Inefficiency caused Loss}\xspace}
\newcommand{\cilfull}{\mbox{Contention Inefficiency caused Loss}\xspace}

\newcommand{\OPNAME}{\mbox{FiCCO}\xspace}
\newcommand{\OPNAMEBOLD}{\mbox{\textbf{FiCCO}}\xspace}

\newcommand{\gemm}{\mbox{GEMM}\xspace}
\newcommand{\gemms}{\mbox{GEMMs}\xspace}

\newcommand{\pyatpLoss}{\mbox{3.9$\times$}\xspace}
\newcommand{\otb}{\mbox{OTB}\xspace}
\newcommand{\mt}{\mbox{MT}\xspace}
\newcommand{\heuAcc}{\mbox{81\%}\xspace}

\newcommand{\mix}{\mbox{AMD Instinct\textsuperscript{\texttrademark} MI300X}\xspace}

\newcommand{\mixp}{\mbox{AMD MI300X Infinity Platform}\xspace}

\def\BibTeX{{\rm B\kern-.05em{\sc i\kern-.025em b}\kern-.08em
    T\kern-.1667em\lower.7ex\hbox{E}\kern-.125emX}}

%% file: 001-abstract.tex
Modern ML workloads demand distributing training and inference across multiple GPUs. 
However, these parallelization techniques often suffer from exposed critical-path communication, leaving a potential 1.7$\times$ speedup on the table through compute-communication overlap.
Prior overlapping methods harness the fact that ML model state and inputs are already sharded into the number of GPUs, and overlap the compute and communication at shard granularity.
However, such coarse-grained overlap suffers from limited network topology support, and  suboptimal dataflows. In this work, we instead make a case for \textit{finer-grain} compute-communication overlap which we term \OPNAMEBOLD. \OPNAME operates one level deeper than traditional sharding, and unlocks overlap for a wider set of network topologies and enables finer-grain dataflow.

We show that \OPNAME opens up a wider design space of execution schedules than possible at shard-level alone. To walk the design space of schedules, we study and characterize the performance inefficiencies on doing overlap and overlay the schedules with the associated inefficiency signatures. 
Our characterization reveals decomposition and contention based slowdowns to be the major performance limiters, and we correlate the slowdown factors with the static compute/communication operator sizes. This helps us design heuristics (that frameworks and runtimes can harness) to select bespoke \OPNAME schedules based on the nature of underlying ML operations. Finally, to further minimize contention inefficiencies inherent with operation overlap, we offload communication to GPU DMA engines. We evaluate several scenarios from realistic ML deployments and demonstrate that our proposed heuristics driven bespoke schedules deliver up to 1.6$\times$ speedup. Further, our heuristics provide accurate guidance to pick the optimal schedule in \heuAcc of unseen scenarios.

%% file: 002-intro.tex
\section{Introduction}
\label{sec:intro}

The rapid escalation of compute and memory demands in machine learning (ML) workloads~\cite{totc} has necessitated large-scale distributed execution. For example, training Llama-3 required nearly 16k GPUs~\cite{scaling-llama-3}. 
These distributed setups employ various parallelization strategies~\cite{gpipe, py_fsdp, cp, gshard} to shard model states and inputs across devices. Such partitioning necessitates frequent communication collectives, such as all-gather, to synchronize and communicate data (e.g., activations) at periodic intervals.

\begin{figure}[!t]
  \centering
  \includegraphics[scale=0.4]{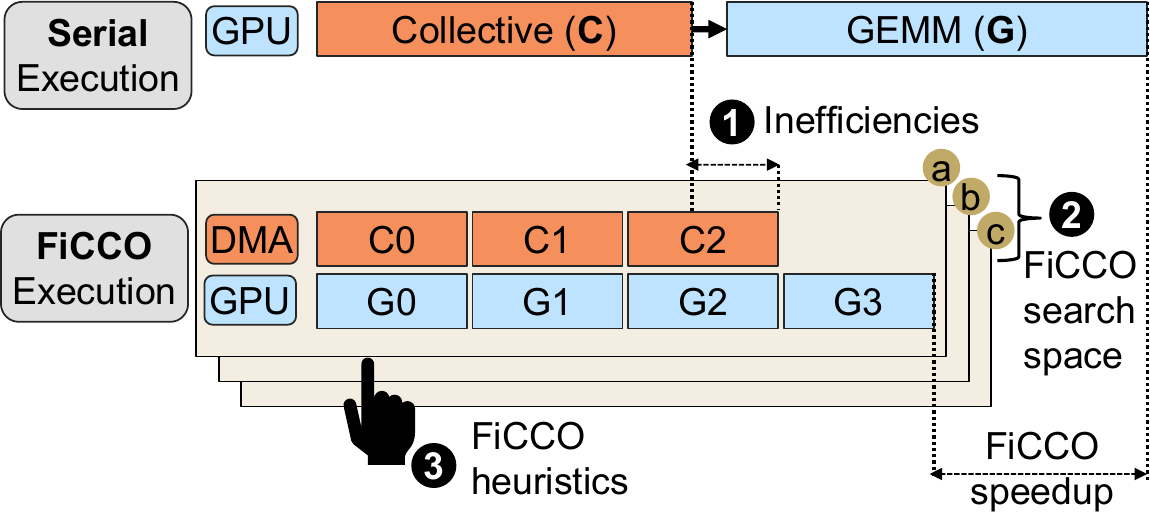}
  \caption{Speedup via finer-grain decomposition of data-dependent communication-computation (\OPNAMEBOLD). \OPNAME heuristics (\ding{174}) picks the optimal schedule within the search space (\ding{173}) to minimize overlap-induced inefficiencies (\ding{172}).}
  \label{fig:intro}
\end{figure}

While independent computation-communication can effectively hide communication latency, data-dependent computation-communication often leaves communication exposed on the critical path. 
An example of the former is fully-sharded data parallelism (FSDP)~\cite{py_fsdp} technique, where weights are partitioned across GPUs and the communication of weights of the next layer can be overlapped with the computation of the current layer. In contrast, examples of data-dependent computation-communication include tensor-sequence parallelism~\cite{seq_parallelism} and context-parallelism~\cite{cp}, wherein communication on critical-path feeds into a data-dependent computation. In such cases, exposed communication provides up to 1.7$\times$ speedup opportunity, that we target in this paper.

\begin{figure*}[!t]
  \centering
  \includegraphics[width=\textwidth]{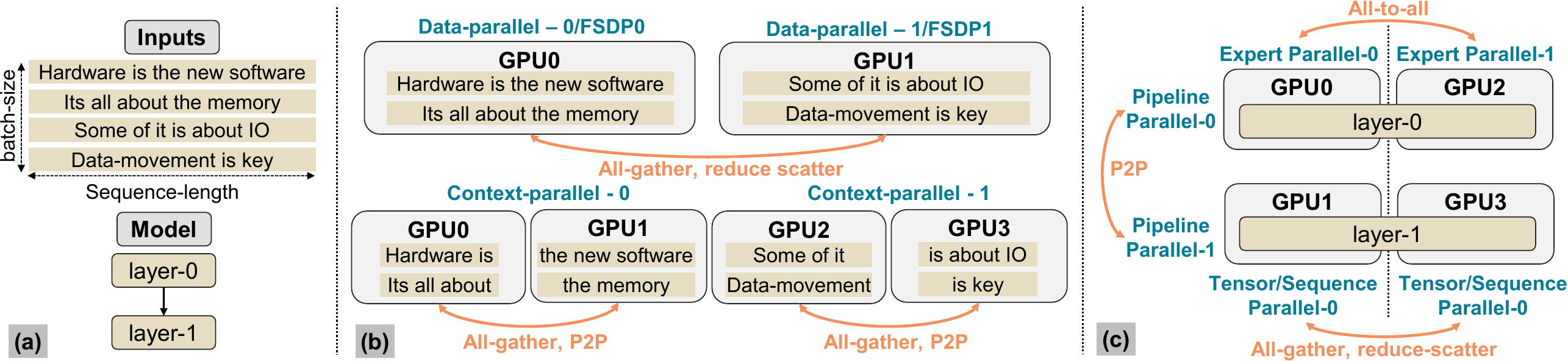}
  \caption{(a) Sample inputs and ML model. (b) Data-parallelism (Fully-Shared Data-Parallel (FSDP)) and context parallelism shard the input. (c) Expert parallelism, context-parallelism, tensor-sequence parallelism shard model state amongst GPUs.}
  \label{fig:bkg_ml_llsm}
\end{figure*}

Prior works~\cite{pyatp,cutlassDist} address these challenges by overlapping computation and communication at the shard granularity, exploiting the inherent sharding in the parallelization techniques (e.g., tensor parallelism shards a single layer's weights equally across GPUs).
However, such coarse-grain shard-based techniques manifest a severe limitation that they harness peer-to-peer communication operations (that is, a GPU communicating with only one other GPU at a time). While suitable for switch-based GPU networks (flexible bandwidth allocation), these techniques leave network links idle with direct-connection based GPU networks, delivering considerably lower performance (up to \pyatpLoss lower). Further, as they inherently operate at shard-granularity they limit granularity of subsequent operations.

We propose \OPNAMEBOLD, a \textit{finer-grained} compute-communication overlap that decomposes communication one level deeper than traditional sharding (for example, in an 8-GPU system, \OPNAME utilizes transfer sizes one-eighth those of shard-based methods). This deeper decomposition enables simultaneous all-to-all communication rather than serial peer-to-peer transfers, thereby reducing latency and maximizing link utilization across diverse topologies. Furthermore, \OPNAME unlocks finer execution granularity for all subsequent operations in the dataflow.

As both shard-based and fine-grain techniques execute smaller compute/communication operators, they lead to various forms of execution inefficiencies (Figure~\ref{fig:intro}), which are highly dependent on the nature of operators being overlapped. That is, the aggregate execution time of decomposed operations can be higher than the execution time of the single larger operation. This happens for myriad reasons including poor utilization and/or contention for GPU resources~\cite{Conccl}. Since execution inefficiencies can vary based on the operator dimensions, a single execution schedule for computation/communication overlap can fail to realize optimal performance. 

To tackle the above challenge, we provide a detailed characterization of inefficiencies incurred as ML operators are decomposed and overlapped, and correlate these inefficiency signatures with the associated operator sizes and shapes. We next explore the \OPNAME design space enabled by finer granularity, which encompasses varying communication shapes, computation uniformity, and computation granularity. The expanded search space of \OPNAME schedules necessitates heuristics that can statically select the optimal schedule. We study how different schedules have tradeoffs in the inefficiency signatures, and correlate them with static operator dimensions to provide heuristics for frameworks and runtimes to pick bespoke \OPNAME schedules.
Finally, we also harness communication offloads to GPU DMA engines to further lower contention based inefficiencies. Our evaluation across several realistic ML deployments demonstrates that our heuristics driven bespoke schedules deliver up to 1.6$\times$ speedup and our heuristics provide accurate guidance to pick the optimal schedule in \heuAcc of unseen scenarios.

In summary, we make the following contributions: 
\begin{itemize}
\item We propose \textit{finer-grain} compute-communication overlap than shard-based, which we term \OPNAMEBOLD, in order to overcome limitations of shard-based overlap. 
\item We provide detailed characterization of inefficiencies incurred as ML operators are decomposed and executed in an overlapped fashion. 
\item We provide a design space of \OPNAME schedules which allows picking bespoke overlap schedules based on the nature of underlying ML operations. Further, we also offload communication to GPU DMA engines to lower contention-driven inefficiencies. 
\item We provide heuristics that guide frameworks and runtimes to pick bespoke \OPNAME schedule for any given scenario, thereby maximizing performance. 
\item We evaluate several scenarios from realistic ML deployments and demonstrate that our proposed bespoke schedules deliver up to 1.6$\times$ speedup and our heuristics picks the optimal schedule in \heuAcc of studied scenarios. 
\end{itemize}

%% file: 003-background.tex
\section{Background}
\label{sec:background}

\begin{figure*}[!t]
  \centering
    \includegraphics[width=\textwidth]{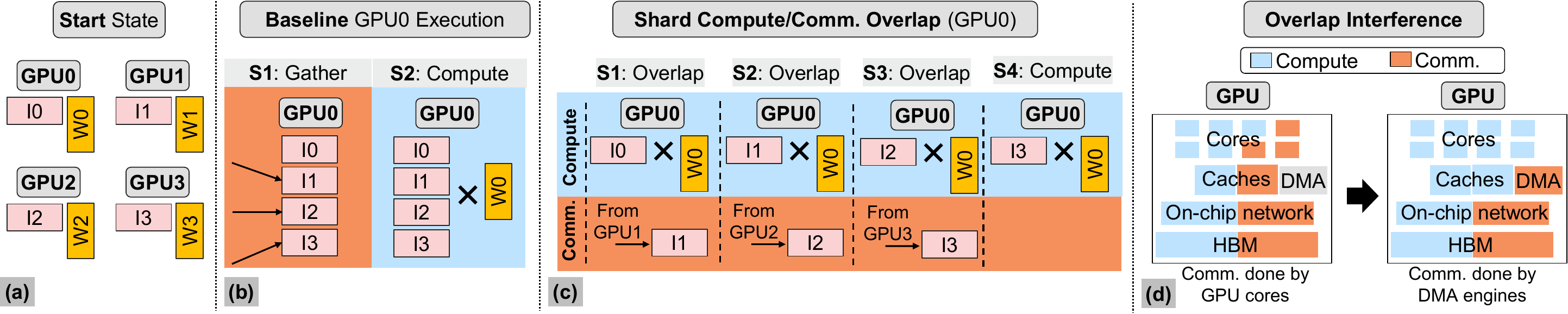}
  \caption{(a) Start state for weights (W) and inputs (I) on a four GPU system. (b) Steps(S) in baseline serial execution of communication and computation.  (c) Shard-based overlap: \textbf{P2P} shard communication and shard-level computation overlap. (d) Contention of limited GPU resources with operation overlap, without and with DMA communication offloads.}
  \label{fig:bkg_shard_overlap_interference}
\end{figure*}

\subsection{ML - Increasingly Distributed}
\label{sec:bkg_ml_dist}
Large language models (LLMs) are becoming ubiquitous, from conversational agents to code generation~\cite{llm-code-gen, llm-code-gen-2,llm-code-gen-3}. The LLM architecture consists of a stack of transformer blocks, each transformer block made up of self-attention \cite{flash_attn} and feedforward (MLP) layers. As LLM model sizes and training dataset sizes scale \cite{llm_mem_capacity,scaling-improves-perf}, the compute and memory throughput of a single GPU is insufficient to meet LLM needs \cite{totc}. Consequently, parallelization techniques shard model state and inputs across multiple GPUs.

Figure~\ref{fig:bkg_ml_llsm} depicts key ML parallelization techniques and their associated communication collectives. Vanilla data-parallelism~\cite{data_parallelism_2,data_parallelism} shards model inputs across GPUs. Fully-sharded data-parallel (FSDP~\cite{py_fsdp}) additionally shards model weights, requiring all-gather (\textbf{AG}) prior to executing associated computation. In addition, reducing scatter (\textbf{RS}) of model gradients, in which gradients across GPUs are summed element-wise, is also necessary (Figure~\ref{fig:bkg_ml_llsm}(b)). Also depicted is context parallelism~\cite{context_parallelism}, which shards single input sequence data across GPUs and can incur either ring-based peer-to-peer (\textbf{P2P})~\cite{ringAttn} or \textbf{AG}~\cite{scaling-llama-3} communication. 

Alternative parallelization techniques shard model state across GPUs (Figure~\ref{fig:bkg_ml_llsm}(c)). In order to scale model parameters without commensurate scaling of compute, mixture-of-expert (MoE) models have gained increasing traction~\cite{hybrid_moe,deepspeed}. Such MoE models often deploy expert parallelism, wherein experts in MLP layers are distributed over multiple GPUs incurring all-to-all (\textbf{AA}) communication to disperse and collect input tokens across experts. Pipeline parallelism~\cite{gpipe}, spreads model layers across GPUs in stages, necessitating \textbf{P2P} communication of outputs/inputs between said stages. Finally, tensor parallelism~\cite{megatron} shards linear layers in transformer block across GPUs, which is often combined with sequence parallelism~\cite{seq_parallelism} to shard sequence dependent operations as well. This necessitates \textbf{AG} and \textbf{RS} collectives across participating GPUs.

\subsection{Shard-based Overlap and Interference}
\label{sec:bkg_shard_overlap}
With wide-spread deployment of distributed ML, effectively hiding resultant inter-GPU communication in the shadow of computation gets increasingly important. This is straightforward in presence of independent computation, such as in FSDP, wherein all-gather of weights for next layer can be done in shadow of computation of the previous layer. However, for other forms of parallelization highlighted in Figure~\ref{fig:bkg_ml_llsm}, this is challenging. As an example, Figure~\ref{fig:bkg_shard_overlap_interference}(a) depicts the start state for tensor-sequence parallel scenarios across four GPUs where each holds a row-sliced input shard (I) and column-sliced weight shard (W). In baseline execution (Figure~\ref{fig:bkg_shard_overlap_interference}b), the computation—a matrix-matrix multiplication (GEMM)—requires every input shard to interact with the weight shard on each GPU. Consequently, an All-Gather collective must first synchronize all input shards across the GPUs before the GEMM kernel is invoked. Section~\ref{sec:eval_shard_limitations} shows that this serialized execution can leave up to 1.7$\times$ of the ideal operator-level performance unrealized.

To overcome this, prior works~\cite{pyatp, cutlassDist} harness the fact that model's inputs/states are already sharded allowing compute-communication overlap as depicted in Figure~\ref{fig:bkg_shard_overlap_interference}(c). Instead of gathering all input shards upfront, shards are sent peer-to-peer between GPU pairs one at a time, while overlapping computation on the previously received shard, thereby attaining compute-communication overlap.

Such overlap of computation and communication is realized on modern GPUs via executing two concurrent GPU kernels, one for computation and the other for communication. Such concurrency, as expected, causes interference between the two as depicted in Figure~\ref{fig:bkg_shard_overlap_interference}(d) across different GPU resources (e.g., cores, caches, HBM, etc.). Prior works~\cite{Conccl, t3, triton-dist, tilelink}, offload communication to existing DMA engines on GPU to lower a substantial portion of this interference. That is, by not using GPU cores to orchestrate communication, compute interference is completely eliminated and a portion of cache interference is eliminated as well. We also harness such DMA offloads. 

%% file: 004-motivation.tex
\section{\OPNAME: Motivation}
\label{sec:motivation}

\subsection{Case for Finer-grain Overlap}
\label{sec:motiv_overview}

While shard-based overlap as employed by prior works is indeed promising, we observe that it suffers from key limitations. To this end, we make a case in this work for \textit{finer-grain} compute-communication overlap which we term \OPNAMEBOLD, wherein communication is decomposed at one-level deeper granularity (that is, transfer sizes one-eighth that of shard-based overlap in an eight GPU system) which allows overcoming of limitations with shard-based overlap (discussed in more detail below). While shard-based techniques execute compute and communicate at shard granularity, we propose to communicate at one-level deeper granularity (additional sharding by number of GPUs) while keeping the computation granularity configurable and tunable (either match or increase granularity than shard-based techniques). 

\begin{figure}[t]
  \centering
  \includegraphics[scale=0.4]{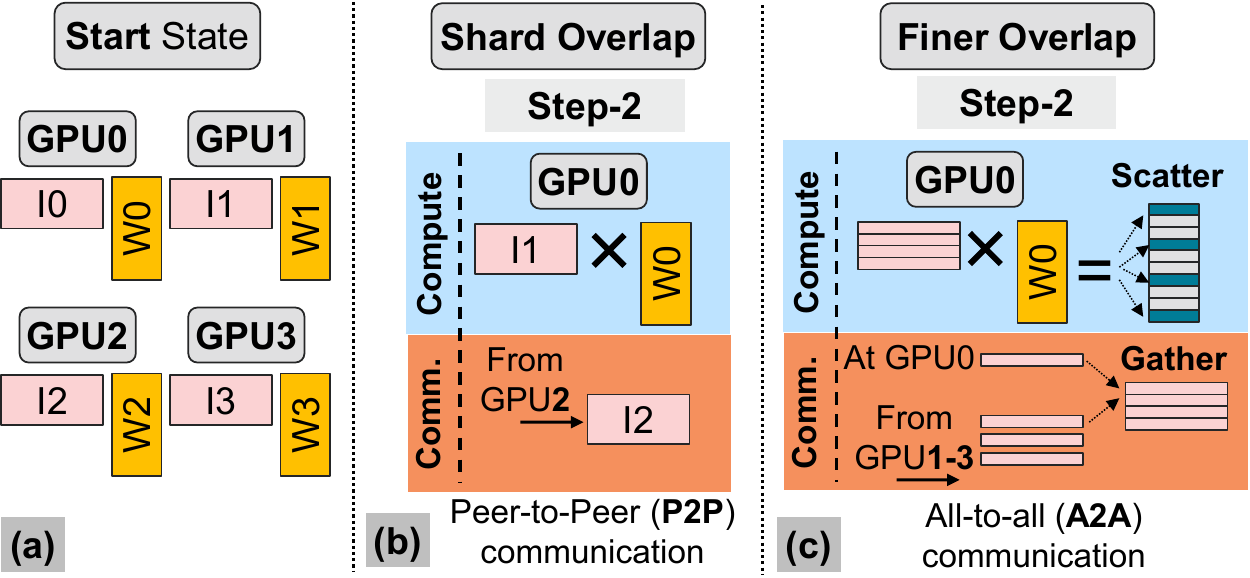}
  \caption{Shared-based overlap versus \OPNAME in action.}
  \label{fig:motiv_finer_overlap}
\end{figure}

\subsection{\OPNAME: Steady State}
\label{sec:motiv_steady_state}
Higher degree of decomposition with \OPNAME unlocks all-to-all communication pattern in steady state as opposed to peer-to-peer communication pattern in shard-based overlap as depicted at a high-level in Figure~\ref{fig:motiv_finer_overlap} ((b) versus (c)). As also depicted, since computation now happens over finer-grain communication received from multiple peer GPUs, depending on the choice of computation granularity (as will be expanded in Section~\ref{sec:ficco_dse}), this induces some additional actions in steady state such as gathering of finer-grain communication buffers (\textbf{Gather}) and potentially scattering of finer-grain outputs in final output space (\textbf{Scatter}). We further discuss in Section~\ref{sec:ficco_dse} how this unlocks a rich design space of \OPNAME schedules.

\subsection{\OPNAME: Benefits}
\label{sec:motiv_benefits}
\begin{figure}[t]
  \centering
  \includegraphics[scale=0.4]{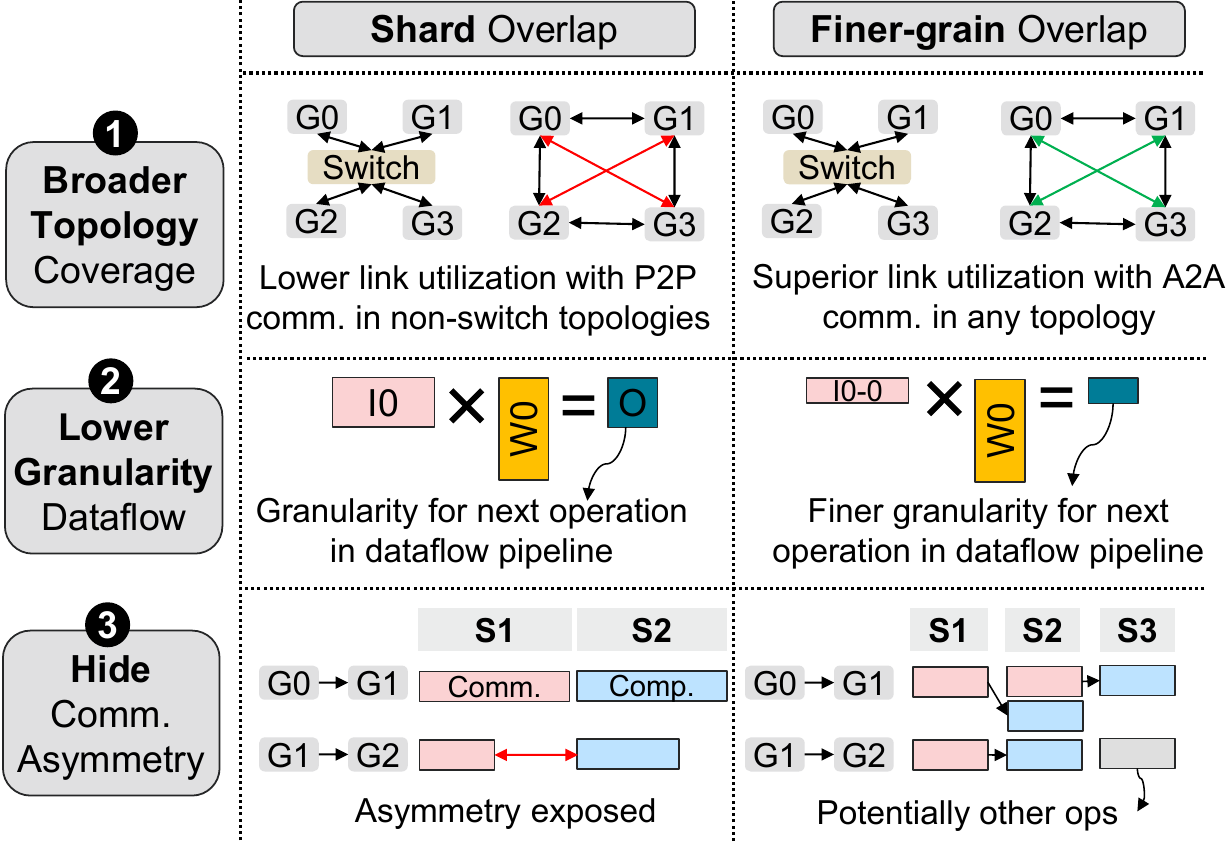}
  \caption{Benefits of Finer-grain overlap. I0,W0: Input, Weight shard in GPU0 respectively; G: GPU; S: Step.}
  \label{fig:motiv_finer_benefits}
\end{figure}

 We depict in Figure~\ref{fig:motiv_finer_benefits} the benefits of \OPNAME in comparison to shard-based overlap. First, \OPNAME broadens compute and communication overlap for data-dependent scenarios for wider topologies. As discussed above, shard-based overlap deploys peer-to-peer communication which is suitable for switch-based GPU topologies ~\cite{nvlink} which allow flexible bandwidth allocation between GPUs. That said, state-of-the-art GPUs such as \mix employ direct connection based topologies. In such cases, peer-to-peer communication can leave network links unused causing performance degradation 
 -- up to \pyatpLoss slowdown compared to serial execution, as discussed further in Section~\ref{sec:eval_shard_limitations}. In contrast, \OPNAME, by decomposing communication one-level deeper (additional sharding by number of GPUs) unlocks all-to-all communication pattern which can keep direct connection based topologies well utilized. Second, finer-grain outputs as enabled by \OPNAME stand to enable lower granularity even in subsequent allowing superior utilization of dataflow-like pipeline. Finally, as also depicted in Figure~\ref{fig:motiv_finer_benefits}, finer-granularity can also allow better hiding of communication asymmetry that can happen with MoE models wherein communication between pair of GPUs can differ based on tokens to be communicated between them. 

%% file: 005-characterization.tex
\section{Data-dependent Compute/Communication Overlap:  Characterization of Inefficiencies}
\label{sec:charac}

As both shard-based or fine-grain techniques execute smaller compute/communication operations as compared to baseline serial execution (Figure~\ref{fig:bkg_shard_overlap_interference}), they lead to various forms of operation execution inefficiencies, which are highly dependent on the nature of operators being overlapped. That is, the aggregated execution time of decomposed operations can be higher than the execution time of the baseline (single large) operation. Careful understanding and characterization of these inefficiencies is necessary in order to effectively manage them, and we provide the same in this section. 

\subsection{Overview of Inefficiency Losses}
\label{sec:charac_overview}
\begin{figure}[t]
  \centering
  \includegraphics[scale=0.4]{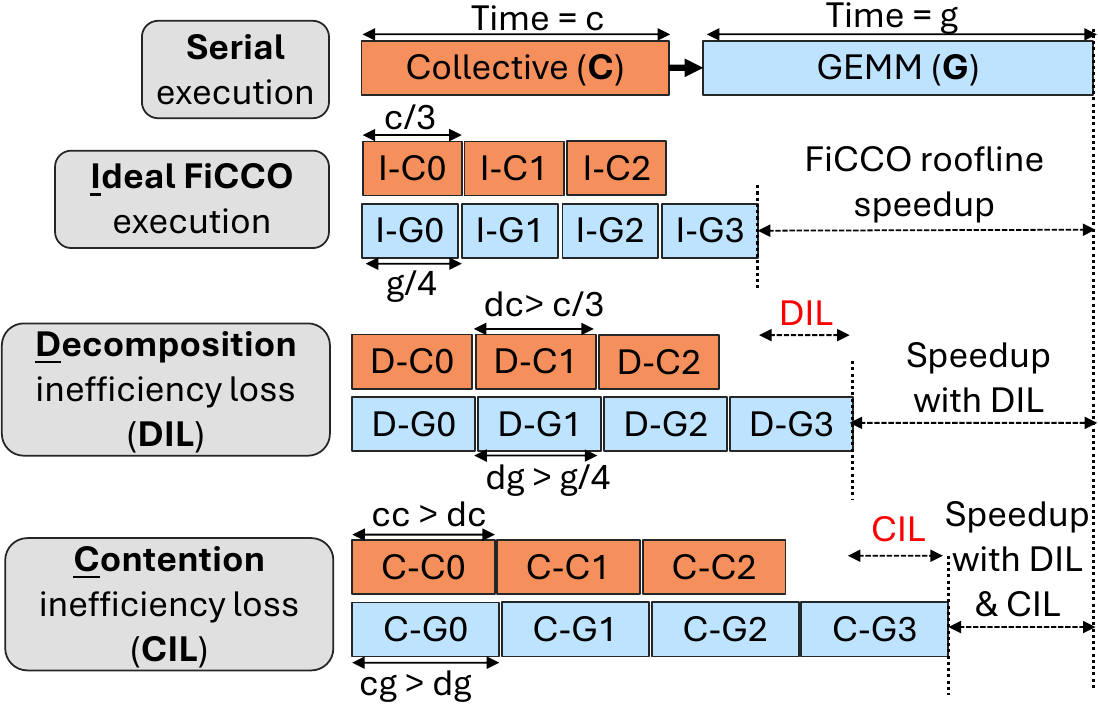}
  \caption{Inefficiencies with operator decomposition and overlap. d[c,g] and c[c,g] represent [Collective,GEMM] times with decomposition and contention respectively.} 
  \label{fig:ficco_dil_cil}
\end{figure}
We depict pictorially in Figure~\ref{fig:ficco_dil_cil} the key inefficiency losses that can incur due to operation decomposition and overlap. Note that the inefficiencies we discuss apply to both shard-based overlap and our proposed \OPNAME overlap. In the characterization sections (Section~\ref{sec:charac_dil}, ~\ref{sec:charac_cil}) we explicitly separate out the empirical differences between the two. 

\textbf{Ideal Execution:} To express the inefficiency losses, we first depict ideal execution in  Figure~\ref{fig:ficco_dil_cil}. In such an execution, decomposing and overlapping of ML operations scales the execution time commensurate to the decomposition degree. This leads to what we refer in this work as \textbf{ideal} speedups for computation/communication overlap. 

\textbf{\dilfull (\dil):} In practice, decomposing operations into smaller sub-operations (as done by both shard-based and \OPNAME techniques) leads to slower execution than predicted by the ideal baseline (Figure~\ref{fig:ficco_dil_cil}). This is because smaller inputs/outputs are less capable of saturating GPU compute units and load-store pipelines, yield poorer cache reuse due to smaller GEMM tile sizes, and suffer from other well-documented inefficiencies~\cite{quantization_ineff}~\cite{cache_util}. We term this as decomposition inefficiency caused loss (DIL henceforth). 

\textbf{\cilfull (\cil):} Additionally, unlike baseline execution where computation and communication run in isolation, both shard-based and \OPNAME techniques overlap these operations, introducing contention in GPU compute and memory subsystems. More specifically, as depicted in Figure~\ref{fig:bkg_shard_overlap_interference}(d), overlapping computation and communication divides GPU compute cores between two concurrent kernels, causing each to receive fewer cores than it would in isolation, an effect we term compute interference. The same phenomenon plays out over caches, network on-chip and HBM. While offloading communication to DMA engines ~\cite{Conccl, t3} 
can eliminate compute interference and some portion of cache interference, memory interference still remains. We term this contention related operation execution slowdown as contention inefficiency caused loss (CIL henceforth). 

\textbf{Other Inefficiency Losses:} We observe in this work that DIL and CIL are the primary inefficiency losses and focus on them. That said, other inefficiency losses exist and we briefly mention them here and omit detailed characterization graphs due to limited space. ML models repeatedly execute the same operation over multiple identical layers across multiple GPUs. We observe variations in execution times of the same operation within the same GPU and across GPUs. However, we find that this variation is within 6\%, so we omit it. Finally, CPU launch of GPU kernels, which increase with both shard-based and our proposed \OPNAME techniques can add launch overheads. These overheads can matter when operation sizes are particularly small. However, graph-based launch techniques can mitigate these overheads~\cite{hipgraph}. 

\input{007-methodology}

\begin{figure*}[t]
    \centering
     \includegraphics[scale=0.4]{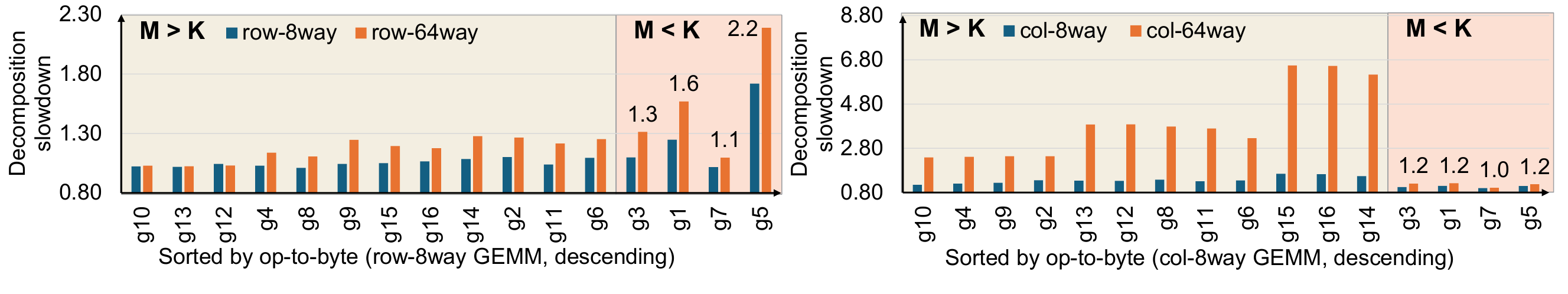}
    \caption{\dilfull (\dil) for GEMM with 8-way and 64-way row (M) or column (K) sharding.}
    \label{fig:ficco_gemm_dil}
\end{figure*}

\subsection{\dil for Compute and Communication}
\label{sec:charac_dil}

\subsubsection{GEMM DIL}
\label{sec:charac_dil_gemm}

We characterize the decomposition overhead as the GEMM is decomposed for shard-based and \textit{finer-grain} overlap. Thus, to evaluate GEMM DIL, we compare the isolated execution of baseline GEMM to that of eight 8-way sharded and sixty-four 64-way sharded smaller GEMMs. As vanilla GEMM is comprised of \textbf{M} rows, \textbf{K} inner reduction columns dimension and \textbf{N} output columns, sharding of activations/gradients can be done in either the row (M) or column (K) dimension. We exercise both. Additionally, note that column-sharding necessitates accumulative GEMM kernels (that is, C += A * B). Finally, note that, 8-way sharding is employed in both shard-overlap and \OPNAME, while 64-way sharding can be employed optionally in \OPNAME. That is, \OPNAME shards communication one-level deeper than shard-based techniques but only optionally shards computation (Section~\ref{sec:ficco_dse}). 

\textbf{Observations:} Figure~\ref{fig:ficco_gemm_dil} depicts DIL for GEMMs. First, as expected, 64-way sharding leads to higher DIL (slowdown) as compared to 8-way sharding for both row and column sharding. Second, DIL depends on relative magnitude of GEMM rows (M) and inner-dimension (K), i.e., row-sharding incurs higher overhead when M \textless~K, and column-sharding when M \textgreater~K. That is, for \textit{g3, g1, g7, g5}, DIL for row-sharding is higher than column-sharding (as depicted with data labels) and vice-versa for rest of GEMMs. Additionally, DIL generally increases as static op-to-byte (\textbf{OTB}) for GEMM (arithmetic intensity, calculated using resultant GEMM dimensions, derived from MNK for ops and bytes) decreases. That is, lower the \otb ratio, more the sensitivity to DIL. Finally, as static GEMM parameters (\otb) dictate DIL, we note this can aid in design of heuristics for picking optimal schedules (Section~\ref{sec:dse_heuristics}).

\subsubsection{Communication DIL}
\label{sec:charac_dil_comm}
\begin{figure}[t]
  \centering
  \includegraphics[scale=0.45]{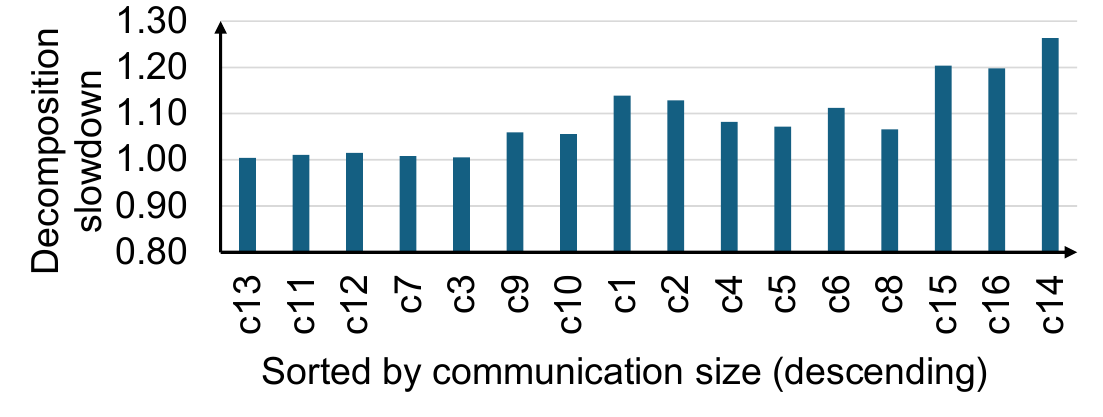}
  \caption{DIL for DMA based all-gather.}
  \label{fig:ficco_comm_dil}
\end{figure}

For communication \dil, as individual transfer size (between a pair of GPUs) in shard-overlap is same as baseline serial execution, shard-overlap does not experience DIL. However, as \OPNAME communicates at finer-granularity, we scale down collective size 8-way for eight GPU system and study resultant slowdown. 

\textbf{Observations:} We depict \dil for communication all-gather in Figure~\ref{fig:ficco_comm_dil}. We observe that communication \dil for \OPNAME has a geomean of about 10\% and positively correlates with communication size as expected. That is, as size increases, there is increased resiliency to DIL as communication gets increasingly bandwidth-bound.

\subsection{\cil for Compute and Communication}
\label{sec:charac_cil}
We compare overlapped and isolated execution of communication/computation to quantify contention overheads.

\begin{figure*}[t]
    \centering
     \includegraphics[scale=0.43]{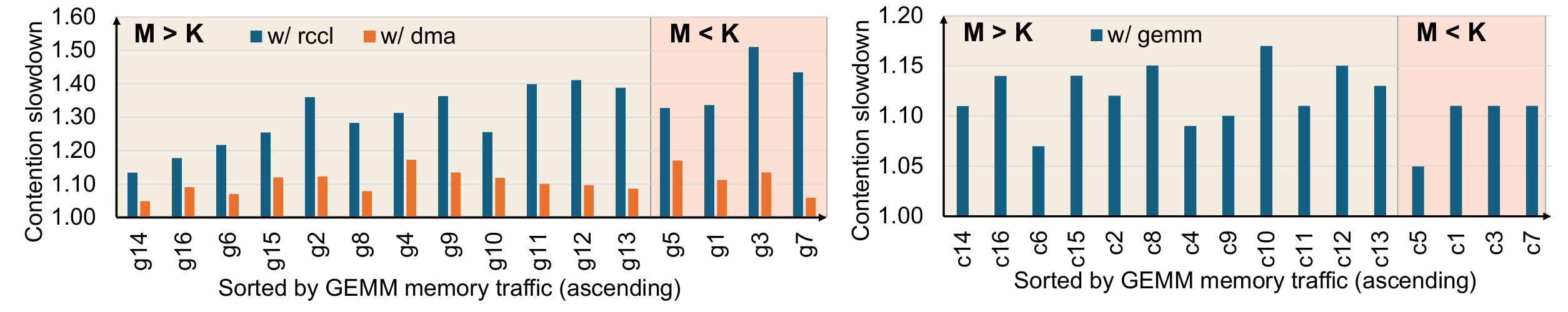}
    \caption{\cilfull (\cil) for GEMM (left) and all-gather communication (right). We compare \cil for GEMM for both RCCL and DMA-based all-gather. \cil for all-gather is in comparison to 8-way M-sharded GEMM.}
    \label{fig:ficco_cil}
\end{figure*}
\subsubsection{GEMM \cil}
\label{sec:charac_cil_gemm}
To evaluate GEMM \cil, we execute 8-way M-sharded GEMM concurrently with RCCL~\cite{rccl} or DMA-based all-gather and report the slowdown relative to isolated execution. In this case, as DIL is already baked into GEMM execution, slowdown observed is purely attributed to contention between GEMM and communication. 

\textbf{Observations:} We depict \cil for GEMM in Figure~\ref{fig:ficco_cil} (left). We first observe that, DMA-based communication causes far lower CIL than traditional GPU core-driven communication (RCCL) in all cases. This is expected as DMAs eliminate compute interference and reduce memory interference~\cite{Conccl}. While clean correlations as in DIL are not as evident for CIL, we observe that CIL generally increases as static memory traffic (\textbf{\mt}) for GEMM increases ( calculated as \textit{MK + NK + MN} for a MNK GEMM). This is so as memory sub-system interference increases with larger GEMM memory traffic. We also evaluate the CIL for shard-based overlap (omitted for space reasons) and observe a geomean \cil of 1.07$\times$ slowdown as opposed to geomean \cil of 1.11$\times$ with \OPNAME. 

\subsubsection{Communication CIL}
\label{sec:charac_cil_comm}
Using the same setup as above we report slowdown in the communication kernel. 

\textbf{Observations:} Figure~\ref{fig:ficco_cil} (right) depicts that communication \cil similarly shows sensitivity to GEMM memory traffic and we observe a geomean \cil of 1.12$\times$ slowdown. In contrast, with larger and more efficient communication, shard-based overlap experiences geomean \cil of 1.03$\times$ slowdown (not depicted for space reasons). 

\begin{figure*}[t]
    \centering
     \includegraphics[scale=0.42]{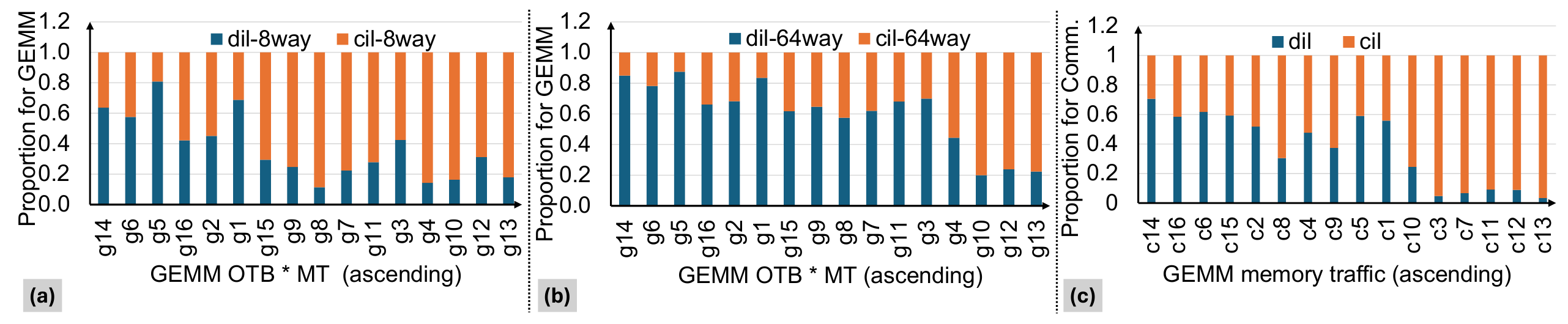}
    \caption{Proportion of DIL versus CIL for GEMMs (g) and all-gather (c).}
    \label{fig:ficco_dil_cil_prop}
\end{figure*}

\subsection{Case for Bespoke \OPNAME Schedules}
\label{sec:charac_bespoke_ficco}
Figure~\ref{fig:ficco_dil_cil_prop} illustrates how DIL and CIL proportionally affect the scenarios under consideration for GEMMs (8-way in (a), 64-way sharding in (b)) and all-gather (c). For GEMMs, as the combination of \otb and \mt increases, GEMMs become more susceptible to CIL (left to right in the figure). This is expected based on trends we discussed above that DIL negatively correlates with \otb and CIL positively correlates with \mt. This effect however, is less prominent for 64-way sharding as the DIL is particularly more prominent there (Section~\ref{sec:charac_dil_gemm}). Communication DIL/CIL proportioning shows similar trend in terms of GEMM \mt. Overall, since different scenarios exhibit different inefficiency signatures, using \OPNAME schedules cognizant of these losses and guided by heuristics for selecting the optimal schedule, stand to deliver better performance.

%% file: 007-methodology.tex
\subsection{Methodology}
\label{sec:methodology}

\begin{table}[t]
  \centering
  \caption{\gemms ~occurring in real world scenarios.}
  \resizebox{\columnwidth}{!}{
  \begin{tabular}{l l l l l}
    \toprule
    Name & Parallelism & Model & \gemm (M,N,K)\\
    \midrule
    g1 & SP+TP    &  llama-3-405b & (16384,16384,131072) \\
    g2 & SP+TP    &  llama-3-405b & (131072,16384,16384) \\
    g3 & SP+TP    &  llama-3-405b & (53248,16384,131072) \\
    g4 & SP+TP    &  llama-3-405b & (131072,53248,16384) \\
    g5 & SP+TP    &  llama-2-70b & (8192,8192,262144) \\
    g6 & SP+TP    &  llama-2-70b & (262144,8192,8192) \\
    g7 & SP+TP    &  llama-2-70b & (28672,8192,262144) \\
    g8 & SP+TP    &  llama-2-70b & (262144,28672,8192) \\
    \midrule
    g9 & SP+TP    &  llama-3-405b & (196608,18432,16384) \\
    g10 & SP+TP    &  llama-3-405b & (196608,106496,16384) \\
    g11 & SP+TP    &  llama-2-70b & (1048576,10240,8192) \\
    g12 & SP+TP    &  llama-2-70b & (1048576,57344,8192) \\
    \midrule
    g13 & EP    &  DeepSeek & (1607680,57344,8192) \\
    g14 & EP    &  Mixtral & (147456,28672,4096) \\
    g15 & EP    &  Mixtral & (327680,28672,4096) \\
    g16 & EP    &  Mixtral & (229376,28672,4096) \\
    \bottomrule
  \end{tabular}
  }
  \label{tab:meth_scenarios}
\end{table}

\subsubsection{System and Execution Setup}
\label{sec:meth_system}
Our setup comprises of \mixp comprised of 8x \mix GPUs \cite{mi300x} with a fully connected topology using AMD Infinity Fabric\textsuperscript{\texttrademark} bi-directional links, each link with a uni-directional bandwidth of 64GB/s. Our setup is based off PyTorch~\cite{pytorch} as it is the most common framework for model training and inference. We focus on overlap of key ML operations: GEMM and communication, ML collectives kernels such as all-gather and all-to-all. For GEMMs, we harness AMD ROCm™ hipblaslt~\cite{hipblaslt} library of high-performance GEMM kernels. For communication, we either harness AMD ROCm communication collectives library (RCCL) ~\cite{rccl}, a standard library of collective communication routines for GPUs or launch memory copies (\textit{hipMemcpyDtoDAsync}~\cite{hip}) to offload communication to GPU DMA engines. We use multiple GPU streams~\cite{hipstream} that allow concurrent execution of GPU kernels. We harness symmetric memory~\cite{symm-mem} for input/output buffers to eliminate intermediate buffer copies and enable direct peer-to-peer GPU memory access. We report the average of five measured runs following ten warm-up iterations.

\subsubsection{Data-dependent Overlap Scenarios Under Consideration}
\label{sec:meth_workload}

We study data-dependent compute-communication scenarios from real-world ML deployments in this work~\cite{scaling-llama-3,llama-2,mlperf_inference,llama,llama-3}, summarized in Table~\ref{tab:meth_scenarios}. More specifically, we study tensor-sequence parallel scenarios wherein, model weights are sharded and activations are all-gathered~\cite{seq_parallelism}. We also study scenarios for expert-parallelism~\cite{deepseek, mixtral}, wherein input tokens are communicated in an all-to-all fashion before executing the expert layers. 

Finally, as we discussed in Section~\ref{sec:bkg_ml_dist}, there exist other data-dependent compute/communication overlap scenarios. We omit some of them that require communication with associated arithmetic operations as GPU DMA engines lack compute capabilities today (e.g., tensor parallelism with reduce-scatter). With future such support, we believe our conclusions/analysis will apply in those scenarios as well. 

%% file: 006-ficco_dse.tex
\section{\OPNAME: Design Space \& Heuristics}
\label{sec:ficco_dse}

We discuss in this section both the design space unlocked via \OPNAME and the schedules we study in this work. 

\subsection{\OPNAME Design Space Dimensions}
\label{sec:dse_dimensions}

Recall from Section~\ref{sec:motivation} that \OPNAME employs \textit{finer-grained} compute-communication overlap, decomposing communication one level deeper than shard-based techniques (for example, in an eight GPU system, one-eighth the transfer size of shard-based overlap). Specifically, while shard-based techniques operate at shard granularity for both communication and computation, \OPNAME decouples the communication and computation granularity. \OPNAME communicates at the finer granularity, while computation granularity is configurable--either matching or finer than shard-based techniques.

\begin{figure*}[t]
     \centering
     \subfloat[]{
		 \includegraphics[scale=0.4]{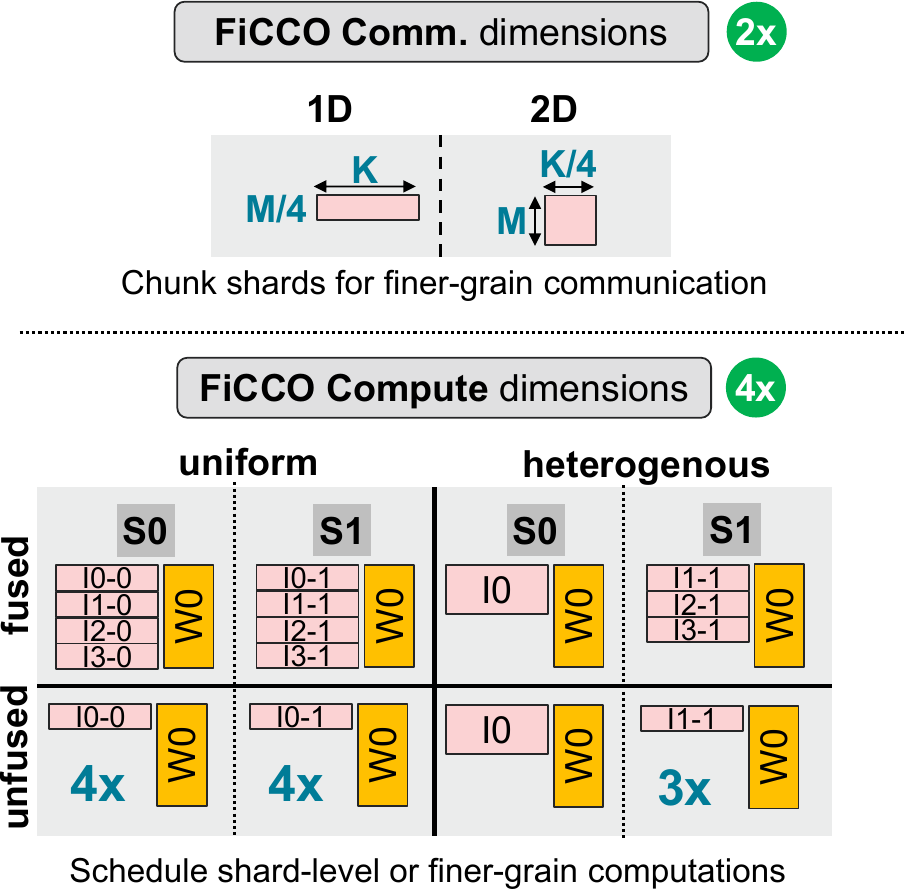}
    	\label{fig:ficco_framework}
     }
          \subfloat[]{
     	\includegraphics[scale=0.4]{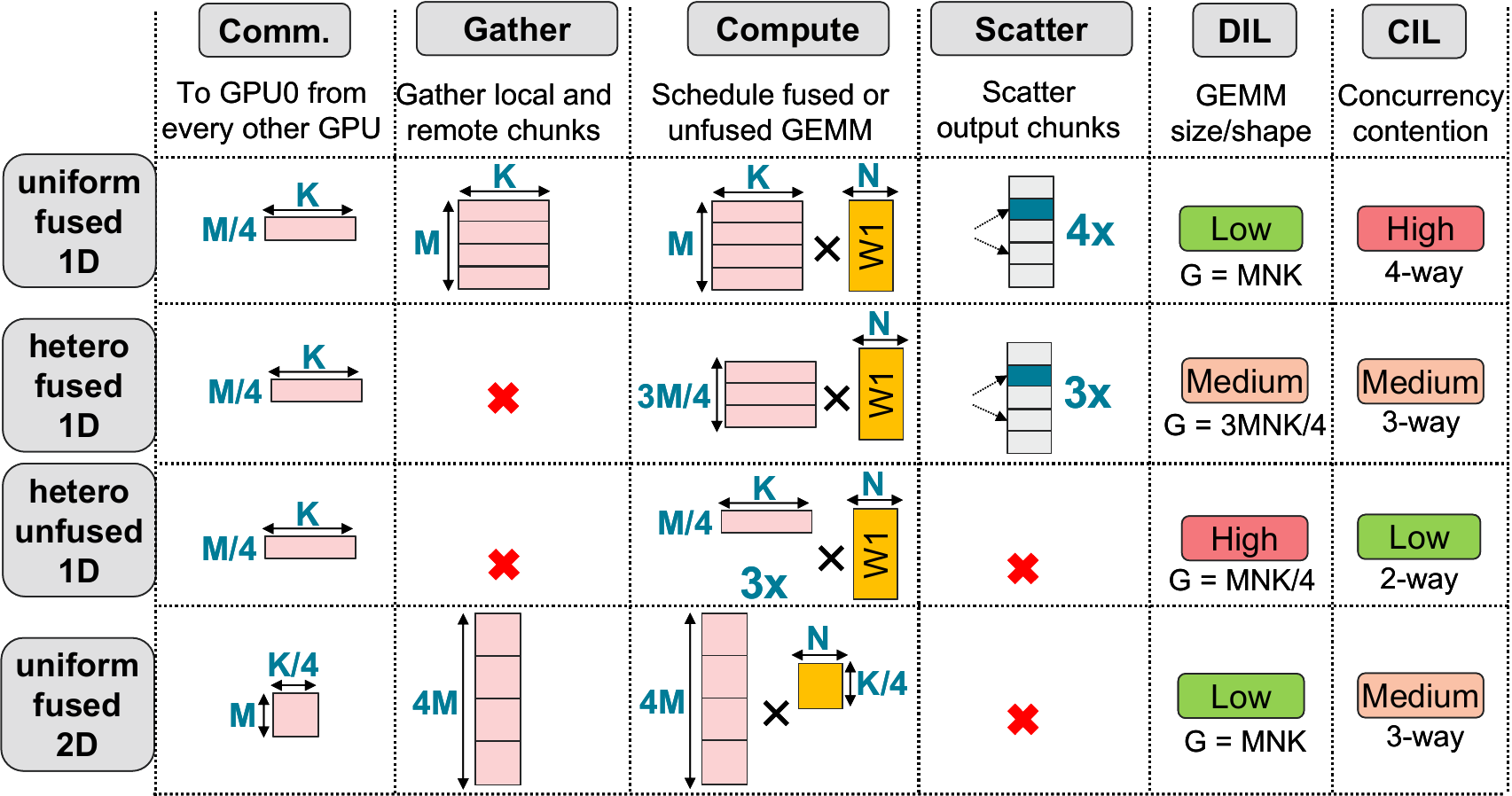}
    	\label{fig:ficco_schedules}
     }
     \caption{(a) \OPNAME design space resulting in eight possible schedules. (S: Step) (b) \OPNAME schedules under consideration.}
\end{figure*}

Consequently, the design space dimensions possible with additional communication sharding in \OPNAME are depicted in Figure~\ref{fig:ficco_framework}. Specifically, for communication, resultant buffers can be 1-dimensional (\textbf{1D}) with sharding in row (M) dimension or 2-dimensional (\textbf{2D}) with sharding in column (K) dimension allowing two choices in this dimension. Further, for computation, \OPNAME unlocks four choices. First, in each overlap step, as GPUs receive buffers from all peer GPUs, either a single/fused GEMM kernel using all the buffers can be executed (\textbf{fused}) or individual kernels allowing flexible scheduling (\textbf{unfused}) can be executed. Second, at the start of the execution (first step), in order to hide exposed communication, each GPU can choose to start executing on its local shard without waiting for any data from peer GPUs leading to heterogeneous subsequent steps (\textbf{hetero}) or combine remote and local buffers to ensure all steps execute the exact same GEMM (\textbf{uniform}). We next discuss concrete \OPNAME schedules to study how each choice manifests tradeoffs in terms of inefficiency loss signatures.

\begin{figure*}[t]
     \centering
    \subfloat[]{
     	\includegraphics[scale=0.4]{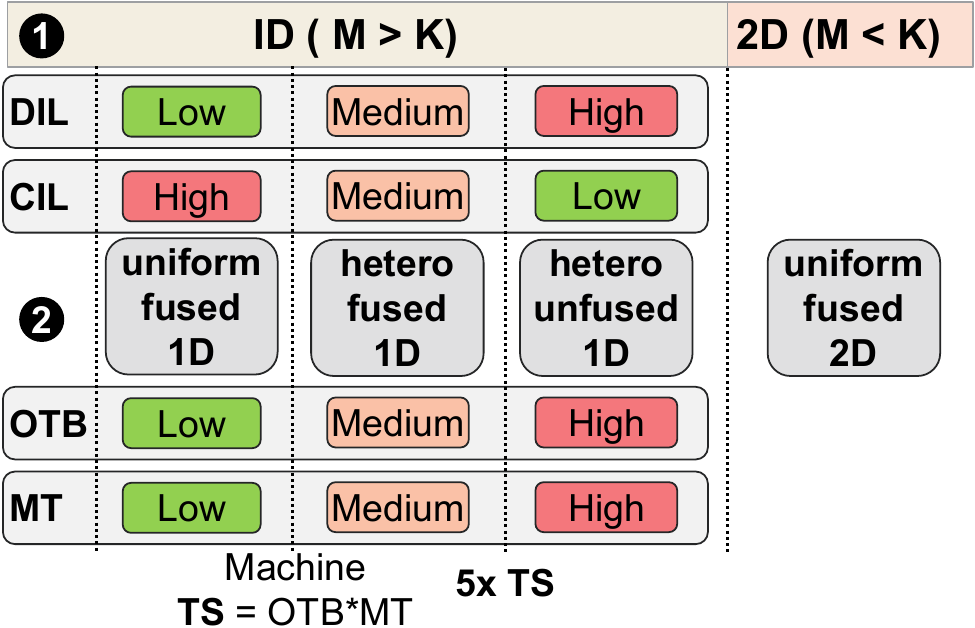}
    	\label{fig:ficco_heuristics}
     }
     \subfloat[]{
		 \includegraphics[scale=0.4, trim=1.8cm 0cm 0.5cm 0cm, clip]{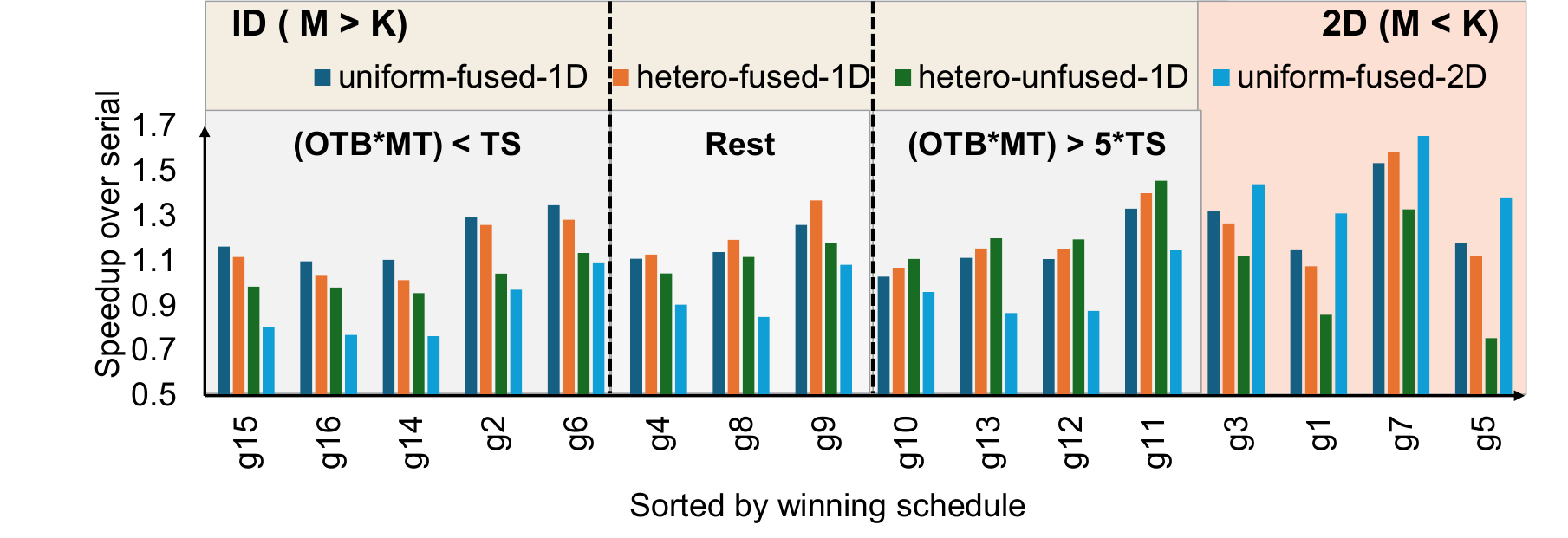}
    	\label{fig:ficco_schedules_eval}
     }
     \caption{(a) \OPNAME heuristics -- GEMM \underline{o}p-\underline{t}o-\underline{b}yte/OTB, GEMM \underline{m}emory \underline{t}raffic/MT). (b) \OPNAME schedules performance.}
\end{figure*}

\subsection{\OPNAME Design Space Schedules}
\label{sec:dse_schedules}
Figure~\ref{fig:ficco_schedules} illustrates the four \OPNAME schedules studied, depicting their steady-state actions. We classify schedules by computation uniformity (\textit{uniform, hetero}), granularity (\textit{fused, unfused}), and communication shape (\textit{1D, 2D}). Also, recall from Section~\ref{sec:motiv_steady_state} that \OPNAME may require gathering of finer-grained communication buffers (\textbf{Gather}) and potentially scattering outputs into the final output space (\textbf{Scatter}), depending on the compute dimension and communication sharding choices. Finally, we assign each schedule an inefficiency loss signature in terms of DIL and CIL, based on its resultant GEMM dimensions and concurrency degree. 

Decoding the provided schedules in a comparative fashion, we observe that all schedules communicate the same effective buffer size and that \texttt{uniform-fused-2D} communicates 2D buffers. Next, to ensure uniformity in GEMM sizes, all uniform schedules incur gather of local and remote received buffers. Next, some schedules either invoke single fused GEMM kernel or some invoke multiple GEMM kernels (unfused). Finally, some schedules scatter output in final output space as they compute on non-contiguous rows in input buffer. We assign DIL degree to schedules based largely on resultant GEMM size for it dictates the GEMM \otb and hence DIL (Section~\ref{sec:charac_bespoke_ficco}). Also, we assign CIL degree to a schedule based on the concurrency degree it manifests (e.g., \texttt{uniform-fused-1D} can execute communication, gather, compute, and scatter at same time), stressing memory traffic which dictates CIL (Section~\ref{sec:charac_bespoke_ficco}).

While with our proposed design space (Section~\ref{sec:dse_dimensions}) a total of eight schedules are possible, we study four of these. This is because we observe that the inefficiency loss signatures of these missing schedules are strictly worse than the ones we study. For the three other 2D schedules (M\textless K), row-sharding is suboptimal. Thus, creating \textit{hetero} or \textit{unfused} schedules that shard in row-dimension would further degrade performance.  Similarly, \texttt{uniform-unfused-1D} further shards \texttt{hetero-unfused-1D}, thereby increasing DIL while keeping CIL unchanged since the Scatter/Gather operations remain unchanged.

\subsection{\OPNAME Heuristics}
\label{sec:dse_heuristics}

Since \OPNAME leads to distinct schedules, selecting the optimal one is essential for maximizing performance. While we can benchmark common GEMM dimensions in distributed ML, the vast diversity of batch sizes, sequence lengths, and model scales makes exhaustive offline profiling infeasible. We therefore utilize heuristics to enable optimal \OPNAME execution based on static parameters, that also works for future/unseen shapes and sizes. The heuristics can aide frameworks and runtimes to pick bespoke schedules.
We present our design for this heuristic in Figure~\ref{fig:ficco_heuristics}. Specifically, we first observe that relative magnitude of GEMM row (M) and column dimension (K) can provide a clear guidance on choice of communication shape to pick: \textbf{1D} if M\textgreater K or \textbf{2D} otherwise in order to minimize resultant DIL (Section~\ref{sec:charac_dil_gemm}). For latter, there is a single \OPNAME schedule available (\textit{uniform-fused-2D}). For the former, first recall that each \OPNAME schedule has a unique and inherent inefficiency loss signature (DIL-CIL degree) as depicted in Figure~\ref{fig:ficco_heuristics}. As such, we match each schedule with operations least sensitive to their inherent inefficiencies. As an example, since \textit{uniform-fused-1D} inherently manifests low DIL and high CIL, it is selected for scenarios with low GEMM op-to-byte (\textbf{\otb}) and low memory traffic (\textbf{\mt}), since DIL negatively correlates with \otb and CIL positively correlates with \mt (Section~\ref{sec:charac_bespoke_ficco}). We similarly match \textit{hetero-unfused-1D} with high combination of \otb and \mt and assign \textit{hetero-fused-1D} schedule otherwise. 

To demarcate the resulting \otb and \mt tranches, we leverage machine-level hardware characteristics. We define a combined machine threshold using combined  \otb and \mt for underlying GPU in terms of its peak compute FLOPs (op-to-byte $\times$ memory bandwidth = FLOPs). We then assign \textit{uniform-fused-1D} to scenarios where the combined \otb and \mt is below this machine-level threshold, while \textit{hetero-unfused-1D} is assigned when the combined metric exceeds the threshold by 5$\times$. We evaluate this heuristic both for scenarios under consideration and for synthetic scenarios and report efficacy in Section~\ref{sec:eval_heuristic}.

%% file: 008-eval.tex
\section{\OPNAME Evaluation}
\label{sec:eval}

\subsection{\OPNAME Realization}
\label{sec:eval_realization}
Our system setup is described in Section ~\ref{sec:meth_system}. \OPNAME schedules are implemented in PyTorch by invoking standard GEMM libraries and DMA transfers, without modifying the underlying optimized kernels. We use \textit{hipStreamWrite} and \textit{hipStreamWait}~\cite{stream_wr_wt} for lightweight synchronization.
To incorporate \OPNAME, the user provides only the GEMM inputs; based on the GEMM dimensions our heuristic will select and execute the optimum overlap schedule, replacing the serial communication and computation.

\subsection{Shard-based Overlap: Limitations}
\begin{figure}[t]
  \centering
  \includegraphics[scale=0.42]{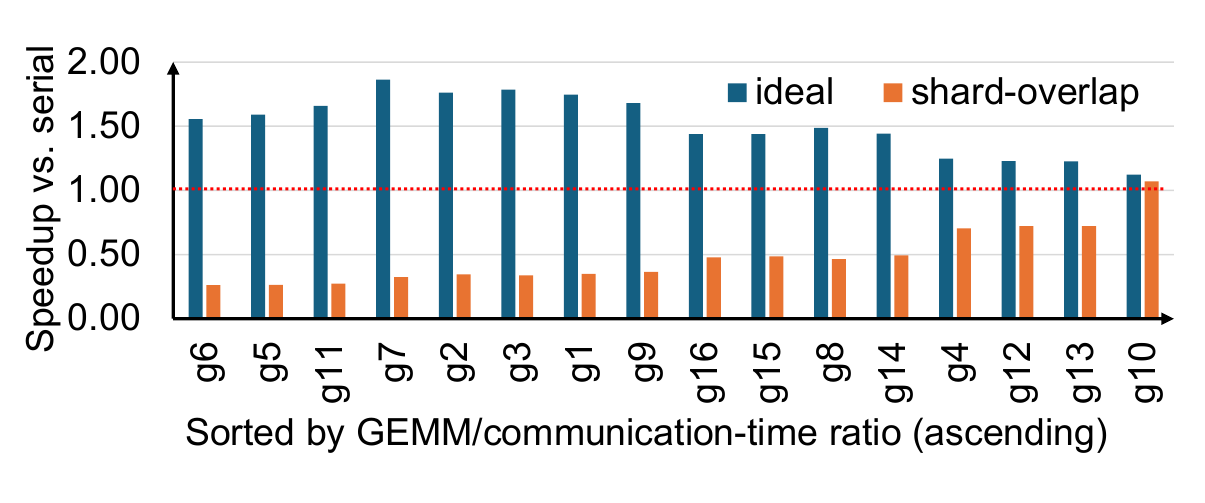}
  \caption{Deficiencies of shard-based overlap.}
  \label{fig:ficco_shard_overlap_eval}
\end{figure}

We start with motivating \OPNAME by demonstrating speedup potential with ideal computation-communication overlap, and limitations of current shard-overlap based techniques in Figure~\ref{fig:ficco_shard_overlap_eval}. Ideal performance assumes operator decomposition scales linearly with no slowdowns from decomposition or contention. We implement shard-overlap following PyTorch Async Tensor Parallelism~\cite{pyatp} technique. 

\label{sec:eval_shard_limitations}

Ideal performance follows a bell curve relative to the GEMM/communication time ratio (along x-axis), that is, the more balanced the operators the greater the benefit of perfect overlap.
However, shard-based overlap, with peer-to-peer communication, under-utilizes the available network links in \mix (we observe 7$\times$ communication slowdown). We observe a negative correlation between speedup and the GEMM/communication time ratio; as GEMM execution time increases (moving rightward on the x-axis), they better mask communication inefficiencies. Regardless, shard-overlap does not attain speedups.

\subsection{\OPNAME Schedules Comparison}
\label{sec:eval_schedule_comp}
Figure~\ref{fig:ficco_schedules_eval} depicts the speedups attained by the \OPNAME schedules we discussed, with our heuristic selections overlaid. Since 2D DMA memory copies are currently unsupported, we emulate them using equal-sized 1D copies. Unlike shard-based methods, \OPNAME leverages all-to-all communication to improve network utilization, attaining speedups of up to 1.6$\times$ with 1D schedules and 1.7$\times$ with emulated 2D schedules.

\subsection{\OPNAME Heuristic Evaluation}
\label{sec:eval_heuristic}
Our proposed heuristic in Figure~\ref{fig:ficco_heuristics} predicts the right schedule for all scenarios we study in this work. To further evaluate the efficacy of this heuristic, we generate sixteen additional synthetic scenarios with diverse OTB and MT combinations. Across these scenarios, our proposed heuristic predicts the right schedule for \heuAcc of scenarios, further underscoring its efficacy. For the scenarios where we mispredict, the schedule proposed by our heuristic loses approximately 14\% of the optimal speedup. 

\subsection{\OPNAME Comparison To Other Overlap Techniques}
\label{sec:eval_other_techniques}
\begin{figure}[t]
  \centering
  \includegraphics[scale=0.4]{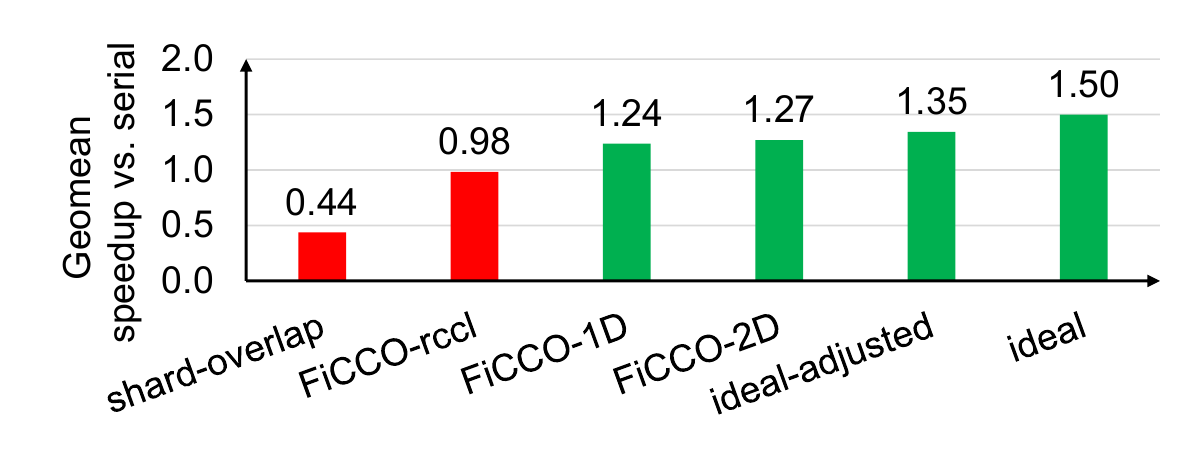}
  \caption{Comparing \OPNAME to other techniques.}
  \label{fig:ficco_others_eval}
\end{figure}

Figure~\ref{fig:ficco_others_eval} depicts geomean speedups across all scenarios for shard-overlap, and \OPNAME 1D and emulated 2D schedules.
We also implement \OPNAME with GPU core-based RCCL communication denoted as \textit{\OPNAME-rccl}. Finally, we attempted comparing against Triton Distributed~\cite{triton-dist} but out-of-memory (OOM) errors in Triton Distributed for our GEMMs precluded successful execution.

%% file: 009-related_works.tex
\section{Related Works}
\label{sec:related}

\begin{table*}[t]
\caption{Comparison of \OPNAME with prior compute-communication overlap works.}
\label{tab:rel_works}
\centering
\small
\renewcommand{\arraystretch}{1}
\setlength{\tabcolsep}{4pt}
\begin{tabularx}{\textwidth}{|>{\hsize=1.6\hsize\raggedright\arraybackslash}X|>{\hsize=0.79\hsize\centering}X|>{\hsize=0.39\hsize\centering}X|>{\hsize=0.97\hsize\centering}X|>{\hsize=0.67\hsize\centering}X|>{\hsize=0.78\hsize\centering}X|>{\hsize=0.8\hsize\centering\arraybackslash}X|}
\hline
\textbf{Name} & 
\textbf{Dependent Comp/Comm} & 
\textbf{DMA-based} & 
\textbf{Use Optimized GEMM Libraries} & 
\textbf{Full-Mesh Support} & 
\textbf{Require no HW Changes} & 
\textbf{Design Space Exploration} \\ 
\hline
CoCoNet~\cite{CoCoNet}, Concerto~\cite{concerto}, Google-Decomp~\cite{google_decomp} & \tickmark & \crossmark & \crossmark & \crossmark & \tickmark & \crossmark \\
\hline
Centauri~\cite{centauri},Domino~\cite{domino}, Comet\cite{comet}, FlashOverlap\cite{flashoverlap} & \tickmark & \crossmark & \tickmark & \crossmark & \tickmark & \crossmark \\
\hline
Flux~\cite{flux}, Fused-Ops.~\cite{fusedOperators} & \tickmark & \crossmark & \crossmark & \tickmark & \tickmark & \crossmark \\
\hline
TileLink~\cite{tilelink} & \tickmark & \tickmark & \crossmark & \tickmark & \tickmark & \crossmark \\
\hline
ConCCL~\cite{Conccl} & \crossmark & \tickmark & \tickmark & \tickmark & \tickmark & \crossmark \\
\hline
AsyncTP~\cite{pyatp}, Dist-Gemm~\cite{cutlassDist} & \tickmark & \tickmark & \tickmark & \crossmark & \tickmark & \crossmark \\
\hline
ACE~\cite{ace} & \tickmark & \crossmark & \tickmark & \tickmark & \crossmark & \crossmark \\
\hline
T3~\cite{t3} & \tickmark & \tickmark & \tickmark & \tickmark & \crossmark & \crossmark \\
\hline
\textbf{\OPNAME} & \tickmark & \tickmark & \tickmark & \tickmark & \tickmark & \tickmark \\
\hline
\end{tabularx}
\end{table*}

Table~\ref{tab:rel_works} compares prior compute-communication overlap works.
There has been significant research to overlap independent computation and communication \cite{independent_c3_1,independent_c3_2,independent_c3_3,independent_c3_4,independent_c3_5,independent_c3_6}, but these solutions do not work in dependent computation-communication. 
CoCoNet\cite{CoCoNet}, Google-Decomposition \cite{google_decomp}, Centauri \cite{centauri}, Flux \cite{flux}, Comet \cite{comet}, Domino\cite{domino}, Concerto \cite{concerto}, FlashOverlap \cite{flashoverlap} decompose the operators, and schedule the overlap such that the dependency is satisfied. However, these solutions do not use DMAs, and therefore incur contention between communication \cite{nccl} and computation kernels for the compute resources. We use DMAs for communication, thereby reducing contention. In addition, several of the solutions require writing a customized GEMM kernel to suit their implementation, and do not use the existing highly optimized GEMM kernels\cite{rocblas,hipblaslt}. We make no changes to the existing GEMM kernels to take advantage of their optimizations.

TileLink \cite{tilelink} and Fused-Operators \cite{fusedOperators} employ fused kernels of compute and communication operators. However, fused kernels often require user intervention to implement communication primitives and tiling optimizations. Furthermore, Fused-Operators does not use DMA, thereby increasing compute contention. DMA based compute communication solutions include ConCCL \cite{Conccl}, PyTorch AsyncTP \cite{pyatp} and Distributed-GEMM \cite{cutlassDist}. ConCCL works only if compute and communication operators are independent. PyTorch AsyncTP, and Distributed-GEMM employ shard-overlap, they are optimized for switch based P2P topologies but do not work well on full-mesh topologies. We employ finer-grain overlap that maximally saturate the bandwidth on full-mesh topologies and achieve good performance. 
ACE \cite{ace} and T3 \cite{t3} propose new hardware modules to handle communication to reduce compute contention. They require hardware changes, while ours is a software solution. 

Finally, based on a thorough characterization of overlap inefficiencies, we provide a design space for finer grain overlap and provide heuristics to pick the optimal one. To the best of our knowledge, this design space exploration has not been done before.

%% file: 010-discussion.tex
\section{Discussion}
\label{sec:discussion}

\subsection{\OPNAME Benefits For Switch-based Topologies}
While we study \OPNAME for direct/fully-connected topologies, its benefits carry over to other topologies as well (e.g., switch-connected topologies). As an example, on decomposing a GEMM, the tradeoff (Section ~\ref{sec:charac}) between increased decomposition loss and reduced contention (due to decreased memory-traffic) would persist even in a switch-topology. Specifically, Figure ~\ref{fig:ficco_schedules_eval}/hetero-infused-1D shows computation sharding incurs lower contention without high decomposition loss and this is agnostic of underlying network topology. As such, \OPNAME-like sharding can enable superior dataflow (Figure ~\ref{fig:motiv_finer_benefits}) for all topologies\footnote{We avoid explicit competitive analysis in compliance with our host industry institution's publication policy.}.

\subsection{\OPNAME Power Behavior}
Concurrent kernel execution can utilize GPU power budget more efficiently via utilization of heterogeneous resources (compute, network, etc.). 
Prior work~\cite{CompPow} analyzes concurrency power and shows that power gets re-allocated between compute and memory based on what is stressed. It also shows that concurrent power can be higher than the isolated execution of compute or communication.
However, \OPNAME harnesses DMA communication which is more energy-efficient~\cite{dmalatte}, and thus \OPNAME power drawn will be lower. Finally, overlapping computation and communication stands to mitigate problematic power swings~\cite{powerstabilizationaitraining,ethan_didt}, which is observed in the serial execution of computation and communication.

\subsection{\OPNAME with Device-initiated Communication}
Recent works such as GPU-initiated networking (GIN)~\cite{gin} enable GPUs to directly initiate communication, eliminating CPU coordination overhead and allowing efficient fusion. While beneficial, they also steal GPU core cycles to initiate and manage communication and as such, both GIN and host-initiated communication are judiciously employed. That said, \OPNAME is fully compatible with both device-initiated and host-initiated communication and can be coupled with either. \OPNAME schedules and associated heuristics are mainly governed by static GEMM dimensions (op-to-byte and memory traffic) and as such these heuristics should also carry over (with one-time tuning cost for thresholds) to device-initiated communication.

%% file: 011-conclusion.tex
\section{Conclusion}
\label{sec:conclusion}
We make a case in this work for \textit{finer-grain} compute-communication overlap which we term \OPNAME, where we argue for finer-granularity, one-level deeper sharding than the commonly deployed shard-level overlap, to unlock compute/communication overlap for wider set of network topologies, finer-grain dataflow and more. We show that \OPNAME opens up a wider design space of execution schedules than possible at shard-level alone and couple this with detailed inefficiency loss characterization to provide heuristics to pick bespoke schedules. Across scenarios from real-world ML deployments, we demonstrate \OPNAME delivers up to 1.6$\times$  speedup.